\documentclass[copyright,creativecommons]{eptcs}
\providecommand{\event}{REFINE 2018} % Name of the event you are submitting to
\usepackage{amsmath}
\usepackage{listings}
\usepackage{enumitem}
\newcommand{\defeq}{\mathrel{\widehat{=}}}

\title{Refining Santa: An Exercise in Efficient Synchronization}
\author{Emil Sekerinski \qquad\qquad Shucai Yao
\institute{Department of Computing and Software \\
McMaster University \\ Hamilton, Ontario, Canada}
\email{emil@mcmaster.ca \quad\qquad yaos4@mcmaster.ca}
}
\def\titlerunning{Refining Santa}
\def\authorrunning{E. Sekerinski \& S. Yao}
\begin{document}
\maketitle

\begin{abstract}
The Santa Claus Problem is an intricate exercise for concurrent programming. This paper outlines the refinement steps to develop a highly efficient implementation with concurrent objects, starting from a simple specification. The efficiency of the implementation is compared to those in other languages.
\end{abstract}

\section{Introduction}

In 1994, Trono proposed the Santa Claus Problem as an exercise in concurrent programming~\cite{Trono94ConcurrencyExercise}:
\begin{quote}
    Santa Claus sleeps in his shop up at the North Pole, and can only be wakened by either all nine reindeer being back from their year long vacation on a tropical island, or by some elves who are having some difficulties making the toys. One elf's problem is never serious enough to wake up Santa (otherwise, he may never get any sleep), so, the elves visit Santa in a group of three. When three elves are having their problems solved, any other elves wishing to visit Santa must wait for those elves to return. If Santa wakes up to find three elves waiting at his shop's door, along with the last reindeer having come back from the tropics, Santa has decided that the elves can wait until after Christmas, because it is more important to get his sleigh ready as soon as possible. (It is assumed that the reindeer don't want to leave the tropics, and therefore they stay there until the last possible moment.) The penalty for the last reindeer to arrive is that it must get Santa while the others wait in a warming hut before being harnessed to the sleigh.
\end{quote}
Trono's original solution uses ten semaphores. The problem is indeed intricate: as Ben-Ari argues, Trono's solution assumes that a signalled process executes immediately: otherwise, when all reindeer are signalled to proceed to the sleigh, some reindeer may still not be harnessed while others have already finished delivering the toys~\cite{BenAri98SantaClaus}. A more robust solution would need additional semaphores for barrier synchronization~\cite{Andrews91ConcurrentProgramming}. Ben-Ari argues that the rendezvous construct of Ada is particularly suitable for this problem and compares a solution in Ada with one in Java using monitors. Downey proposes a solution of a simplified problem employing only four semaphores, but makes the assumption that a signalling process does not continue~\cite{Downey16Semaphores}; under some schedulers, e.g. the semaphore implementation of Python, the first elf runs forever.

The Santa Claus Problem follows a line of whimsically named concurrency problems (see \cite{Downey16Semaphores} for a beautiful collection of those) that all are representative for specific aspects: here, these are {\em priority} (the reindeer have priority over elves), {\em multi-party synchronization} (all reindeer have to be present to engage with Santa and Santa engages either with reindeer or elves), {\em barriers} (all reindeer have to be harnessed, then they jointly ride with Santa, then Santa dismisses them), and {\em batch processing} (Santa consults elves one by one, but only if a group of three is present). The Santa Claus Problem has been used to illustrate concurrency constructs, e.g.~\cite{Benton03JingleBells,Buhr16ConcurrencyConstructs,CameronEtAl06SantaClausStateClasses,Dovland06LiberatingCoroutines,Nienaltowski07ContractConcurrentOOP} and for comparing concurrency constructs~\cite{HurtPedsersen08SantaClausComparison}. Peyton Jones gives a solution in Haskell using software transactional memory~\cite{PeytonJones07BeautifulConcurrency}. Welch and Pedersen present a process-oriented solution using Occam and discuss model-checking a CSP formulation of the problem~\cite{WelchPedersen10SantaClausProcess}. %The reader is invited to compare these designs with the one proposed here.

This paper develops a solution using concurrent objects by a series of refinement steps. The thrust is to start the development with a specification that is as simple as possible, to add details about Santa, the reindeer, the elves, and their interaction in refinement steps, and to arrive at an implementation that is comparable to other efficient implementations. This work is part of an ongoing research program in developing a highly efficient implementation~\cite{MooreOlivaSekerinskiYao14MultiThreadedStack,Yao18} of concurrent objects together with an accompanying verification and refinement theory~\cite{Sekerinski05ConcurrentObjects}.

The next section introduces concurrent objects with guard-based synchronization and discusses the assumptions about atomicity. A general refinement rule for concurrent objects is given and informally justified. This is followed by the presentation of the Santa Claus problem, the development of a solution in five refinement steps, the timing results comparing four implementations, and a discussion. The proofs of the refinement steps are sketched but not carried out in full detail: our goal is to  argue  that the chosen model of concurrent objects allows both highly efficient implementations and intuitive correctness reasoning.

\section{Concurrent Objects}

Concurrent objects here consists of {\em fields}, {\em methods}, and {\em actions}~\cite{BosangueKokSere99ObjectBasedDistributed,BuchiSekerinski00RefiningConcurrentObjects,Misra02ObjectBasedMultiprogramming,Sekerinski05ConcurrentObjects}. Methods must be called to execute but an action can execute on its own whenever its {\em guard} is true. Only one method or action can execute at a time in one object, but all objects can execute concurrently. Objects communicate through method calls; no separate mechanism is needed. For synchronization of objects, methods may also have a guard, which can block the caller.  Consider class {\em Santa}:

\begin{lstlisting}
class Santa
    var s: {Sleeping, Working} = Sleeping
    method wakeup()
        s = Sleeping $\to$ s := Working
    action
        s = Working $\to$ s := Sleeping
\end{lstlisting}

\noindent When object {\em st} is created by \lstinline|st := new Santa|, the method {\em wakeup} can be called, \lstinline|st.wakeup()|. The call blocks if field $s$ of $st$ is not equal to {\em Sleeping} and sets $s$ to {\em Working} otherwise. The single action of the object is executed on its own when its guard is true, {\em s $=$ Working}, and then sets field $s$ to {\em Sleeping}. Thus this represents a Santa who needs to be woken up externally, but will go to sleep on his own.

The guards of methods and actions of an object can depend only on fields of that object; the guard cannot refer to fields of other objects or contain calls. This restriction is meant to allow for an efficient implementation: all objects can evaluate their guards concurrently without interference; a guard can change its value by execution with an object, hence guards only need to be reevaluated after a method or action in that object executes.

All methods and actions are executed atomically, up to method calls. For example, if $S$ is a statement without calls, the sequence \lstinline|st.wakeup()  ; S ; st.wakeup()| executes the first call \lstinline|st.wakeup()| atomically, then $S$ atomically, then the second call \lstinline|st.wakeup()| atomically. Using angular brackets to denote atomic regions, this is equivalent to \lstinline|$\langle$ st.wakeup() $\rangle$; $\langle$ S $\rangle$ ; $\langle$ st.wakeup() $\rangle$|. Both calls to {\em wakeup} may block and delay execution until the guard holds, i.e. Santa is sleeping again. In general, if the execution of a method or action is suspended, another method or action may start to execute or continue execution. There can be arbitrarily many suspended method executions in an object. Once an action is chosen, that action will be executed until termination before another action can be initiated, hence at most one action execution can be suspended. There can only be as many concurrent executions as there are objects.

\newtheorem{theorem}{Rule}
\newcommand{\Skip}{\textit{skip}}
\newcommand{\Class}{\textbf{class }}
\newcommand{\Var}{\textbf{var }}
\newcommand{\Method}{\textbf{method }}
\newcommand{\Action}{\textbf{action }}

We assume that all fields of an object are {\em private} to the object, i.e.~are accessed only by the methods and actions of the object. The sole purpose of classes is to create objects. In general, class $D$ refines class $C$ if a $D$ object can be used instead of a $C$ object. The following rule formalizes refinement with a {\em coupling relation} that relates the fields of $C$ and~$D$:

\begin{theorem}[Class Refinement] Consider classes $C, D$ with list $f, g$ of fields, initialized to $f_0, g_0$, with methods $m_k()$ with bodies $S_k, T_k$, and with actions $A_i, B_j$. Class $D$ may have new methods $n_l()$ with bodies~$U_l$:
\setlength{\jot}{0pt}\rm
\begin{align*}
    &\Class C         &&\textbf{class } D\\
    &~~~~\textbf{var } f = f_0 &&~~~~\textbf{var } g = g_0\\
    &~~~~\textbf{method } m_k()  &&~~~~\textbf{method } m_k() \\
    &~~~~~~~~S_k                 &&~~~~~~~~T_k \\
    &~~~~\textbf{action } A_i    &&~~~~\textbf{method } n_l() \\
    &                          &&~~~~~~~~U_l \\
    &                          &&~~~~\textbf{action } B_j
\end{align*}\it
Class $C$ is refined by class $D$ if for some relation $R$ over $f$ and $g$:
\begin{itemize}
    \item[\rm ($\bf I$)] $R\,f_0\,g_0$
    \item[\rm ($\bf M$)]  $S_k \sqsubseteq_R T_k$ \it ~~for all $k$
    \item[\rm ($\bf N$)] $\Skip \sqsubseteq_R U_l$ ~~or~~ $A_i \sqsubseteq_R U_l$ \it ~~for all $l$ and some $i$
    \item[\rm ($\bf A$)]  $A_i \sqsubseteq_R B_j$ \it ~~for all $j$ and some $i$
\end{itemize}
\end{theorem}

\noindent Condition ($\bf I$) requires that the field initializations have to establish the coupling relation $R$. Condition~($\bf M$) requires that each method of $D$ refines the corresponding method of $C$ through $R$. Condition ($\bf M$) requires that new methods of $D$ either {\em stutter}, i.e.~refine $\Skip$ through $R$, or refine some action of $C$ through~$R$. Condition ($\bf A$) requires that all actions of $D$ refine some action of $C$. Note that not all actions of $C$ have to be refined, i.e.~$D$ can restrict the behaviour of $C$.

For the refinement of statements through a relation, we give only a single rule and appeal to intuition otherwise:

\begin{theorem}[Guarded Assignment Refinement] Let $b$, $d$ be expressions over variables $x$, let $c$, $e$ be expressions over variables $y$, and let $R$ be a relation between $x$ and $y$:
\begin{itemize}
    \item[] $b \to x := d \sqsubseteq_R c \to y := e $ ~~if~~ $R\,x\,y \land c \Rightarrow b$ ~~and~~ $R\,x\,y \land c \Rightarrow R\,d\,e$
\end{itemize}
\end{theorem}

\noindent In refinement steps, new classes may be introduced and objects of those classes may be created. Above rules are applied to ensure that the behaviour of existing objects is preserved.

\section{Refining Santa}

In the development below, subscripts are used to distinguish names across refinement steps.

\subsection*{Specification: Santa's Cycle}

The activity at the North Pole centers around Santa. In the simplest form, Santa either sleeps or works. This is expressed by a class with one field for Santa's state and two actions that switch between these two states, whenever Santa feels like doing so:

\begin{lstlisting}
class Santa$_0$
    var s: {Sleeping, Working} = Sleeping
    action s = Sleeping $\to$ s := Working
    action s = Working $\to$ s := Sleeping
\end{lstlisting}

\noindent A single object {\em st} of class {\em Santa$_0$} is created:

\begin{lstlisting}
st := new Santa$_0$
\end{lstlisting}

\subsection*{Refinement 1: Splitting Santa's Work}

Santa's work consists of either delivering toys or helping the elves: when Santa wakes up, he may either go to state {\em Delivering} or {\em Helping}:

\begin{lstlisting}
class Santa$_1$
    var s: {Sleeping, Delivering, Helping} = Sleeping
    action s = Sleeping $\to$ s := Delivering
    action s = Sleeping $\to$ s := Helping
    action s = Delivering $\to$ s := Sleeping
    action s = Helping $\to$ s := Sleeping
\end{lstlisting}

\newcommand{\Santa}{\textit{Santa}}
\newcommand{\Working}{\textit{Working}}
\newcommand{\Sleeping}{\textit{Sleeping}}
\newcommand{\Delivering}{\textit{Delivering}}
\newcommand{\Helping}{\textit{Helping}}

\noindent A single object {\em st} of class $\Santa_1$ is created:

\begin{lstlisting}
st := new Santa$_1$
\end{lstlisting}

\noindent For applying the rule for Class Refinement, as the coupling relation between $\Santa_0$ and $\Santa_1$ we take:

\begin{itemize}
    \item[] $R_1\, s_0\, s_1 \defeq s_0 = \Working \equiv s_1 \in \{\Delivering, \Helping\}$
\end{itemize}

\noindent Since there are no methods in $\Santa_1$, refinement follows from the conditions for the initialization and the four actions of $\Santa_1$:
\begin{itemize}
    \item[($\bf I$)] $R_1\,\Sleeping\,\Sleeping$
    \item[($\bf A_1$)]  $s_0 = \Sleeping \to s_0 := \Working \sqsubseteq_{R_1} s_1 = \Sleeping \to s_1 := \Delivering$
    \item[($\bf A_2$)]  $s_0 = \Sleeping \to s_0 := \Working \sqsubseteq_{R_1} s_1 = \Sleeping \to s_1 := \Helping$
    \item[($\bf A_3$)] $s_0 = \Working \to s_0 := \Sleeping \sqsubseteq_{R_1} s_1 = \Delivering \to s_1 := \Sleeping$
    \item[($\bf A_4$)] $s_0 = \Working \to s_0 := \Sleeping \sqsubseteq_{R_1} s_1 = \Helping \to s_1 := \Sleeping$
\end{itemize}
Conditions ($\bf A_1$) to ($\bf A_4$) hold by the rule for Guarded Assignment Refinement.

\subsection*{Refinement 2: Introducing Santa's Sleigh}

\newcommand{\Harnessing}{\textit{Harnessing}}
\newcommand{\Riding}{\textit{Riding}}
\newcommand{\Sleigh}{\textit{Sleigh}}
\newcommand{\Back}{\textit{Back}}
\newcommand{\Pulling}{\textit{Pulling}}
\newcommand{\True}{\textit{true}}
\newcommand{\False}{\textit{false}}

Santa's shop coordinates the elves and Santa's sleigh coordinates the reindeer. We first introduce and prioritize the sleigh, postponing the introduction of the reindeer, the shop, and the elves. Here, the sleigh is an active object: the sleigh signals to Santa that all reindeer are back, then Santa harnesses all reindeer, then the reindeer pull the sleigh until Santa releases all reindeer and sleeps again. The synchronization is expressed by the sleigh calling newly introduced methods of $\Santa$. Class $\Santa_2$ splits the $\Delivering$ state of $\Santa_1$ into $\Harnessing$ and $\Riding$; field $b$ is true if the reindeer are back from vacationing:

\begin{lstlisting}
class Santa$_2$
    var s: {Sleeping, Harnessing, Riding, Helping} = Sleeping
    var b: boolean = false
    method back()
        b := true
    method harness()
        s = Harnessing $\to$ s := Riding
    method pull()
        s = Riding $\to$ s, b := Sleeping, false
    action s = Sleeping $\land$ b $\to$ s := Harnessing
    action s = Sleeping $\land \neg$ b $\to$ s := Helping
    action s = Helping $\to$ s := Sleeping

class Sleigh$_2$(st: Santa$_2$)
    var s: {Back, Harnessing, Pulling} = Back
    action s = Back $\to$ s := Harnessing ; st.back()
    action s = Harnessing $\to$ s := Pulling ; st.harness()
    action s = Pulling $\to$ s := Back ; st.pull()
\end{lstlisting}

\noindent Object $st$ of class $\Santa_2$ and object $sl$ of class $\Sleigh_2$ are created; these objects can execute concurrently:

\begin{lstlisting}
st := new Santa$_2$ ; sl := new Sleigh$_2$(st)
\end{lstlisting}

\noindent The first action of $\Sleigh_2$ calls $st.back()$, which executes immediately but under mutual exclusion with any other action of $st$. The calls $st.harness()$ and $st.harness()$ will block until the corresponding guard is true. For applying the rule for Class Refinement to show that $\Santa_2$ refines $\Santa_1$, we take as the coupling relation:

\begin{itemize}
    \item[] $R_2\, s_1\, (s_2, b_2) \defeq
        (s_1 = \Delivering \equiv s_2 \in \{\Harnessing, \Riding\}) \land (s_1 = \Delivering \Rightarrow b_2)$
\end{itemize}
Refinement follows as $back$ and $harness$ of $\Santa_2$ stutter under $R_2$ and $pull$ refines the action $s = \Delivering \to s := \Sleeping$ of $\Santa_1$ under $R_2$. Field $b$ of $\Santa_2$ reduces the nondeterminism that is present among the actions of $\Santa_1$. Formally, the conditions are:
\begin{itemize}
    \item[($\bf I$)] $R_2\,\Sleeping\,(\Sleeping, \False)$
    \item[($\bf N_1$)] $skip \sqsubseteq_{R_2} b := \True$
    \item[($\bf N_2$)] $skip \sqsubseteq_{R_2} s_2 = \Harnessing \to s_2 := \Riding$
    \item[($\bf N_3$)] $s_1 = \Delivering \to s_1 := \Sleeping \sqsubseteq_{R_2} s_2 = \Riding \to s_2, b_2 := \Sleeping, \False$
    \item[($\bf A_1$)]  $s_1 = \Sleeping \to s_1 := \Delivering \sqsubseteq_{R_2} s_2 = \Sleeping \land b_2 \to s_2 := \Harnessing$
    \item[($\bf A_2$)]  $s_1 = \Sleeping \to s_1 := \Helping \sqsubseteq_{R_2} s_2 = \Sleeping \land \neg b_2 \to s_2 := \Helping$
    \item[($\bf A_3$)] $s_1 = \Helping \to s_1 := \Sleeping \sqsubseteq_{R_2} s_2 = \Helping \to s_2 := \Sleeping$
\end{itemize}
These follow from the rule for Guarded Assignment Refinement.

\subsection*{Refinement 3: Introducing Reindeer}

This step leaves Santa unchanged, refines Santa's sleigh into a passive sleigh, and introduces active reindeer. The sleigh coordinates the reindeer by keeping a count, $c$, for the number of reindeer that need to come back, that need to be harnessed, and that need to be pulling. The reindeer cyclically call the {\em back}, {\em harness}, {\em pull} methods of the sleigh. Since reindeer are not further refined, this is simply expressed by a single action composing these calls in sequence rather than by three actions.

\begin{lstlisting}
class Sleigh$_3$(st: Santa$_2$)
    var s: {Back, Harnessing, Pulling} = Back
    var c: 0 .. 9 = 9
    method back()
        s = Back $\to$ c := c - 1 ; if c = 0 then (s, c := Harnessing, 9 ; st.back())
    method harness()
        s = Harnessing $\to$ c := c - 1 ; if c = 0 then (s, c := Pulling, 9 ; st.harness())
    method pull()
        s = Pulling $\to$ c := c - 1 ; if c = 0 then (s, c := Back, 9 ; st.pull())

class Reindeer$_3$(sl: Sleigh$_3$)
    action sl.back() ; sl.harness() ; sl.pull()
\end{lstlisting}

\noindent One sleigh and nine reindeer are created:

\begin{lstlisting}
sl := new Sleigh$_3$ ; for i := 1 to 9 do new Reindeer$_3$(sl)
\end{lstlisting}

\noindent As a note, the refinement is also correct is more or fewer than nine reindeer are created: if there are more than nine reindeer, the first nine arriving will be harnessed; if there are fewer than nine reindeer, Santa can only occupy himself with the Elves and no presents will be delivered! The coupling relation between {\em Sleigh$_2$} and {\em Sleigh$_3$} includes the identity relation on $s$ and restricts $c$ to be between $1$ and $9$:

\begin{itemize}
    \item[] $R_3\, s_2\, (s_3, c_3) \defeq s_2 = s_3 \land 1 \leq c_3 \leq 9$
\end{itemize}

\noindent For brevity, we refer to the body of method $m$ of class $C$ as $C.m$. Refinement of {\em Sleigh$_2$} by {\em Sleigh$_3$} then follows from:

\begin{itemize}
    \item[($\bf I$)] $R_2\,Back\,(Back, 9)$
    \item[($\bf N_1$)] $\Skip \sqsubseteq_{R_3} \Sleigh_3.back$ ~or~ $s_2 = \Back \to s_2 := \Harnessing ; st.back()  \sqsubseteq_{R_3} Sleigh_3.back$
    \item[($\bf N_2$)] $\Skip \sqsubseteq_{R_3} \Sleigh_3.harness$ ~or~ $s_2 = \Harnessing \to s_2 := \Pulling ; st.harness() \sqsubseteq_{R_3} Sleigh_3.harness$
    \item[($\bf N_3$)] $\Skip \sqsubseteq_{R_3} \Sleigh_3.pull$ ~or~ $s_2 = \Pulling \to s_2 := \Back ; st.pull() \sqsubseteq_{R_3} \Sleigh_3.pull$
\end{itemize}
To show $(\bf N_1)$, we distinguish the cases $c > 1$ and $c = 1$: if $c > 1$ initially, then $\Sleigh_3.back$ simplifies to $s = \Back \to c := c - 1$, which refines $\Skip$ under $R_3$; if $c = 1$ initially, then $\Sleigh_3.back$ simplifies to $s = \Back \to s, c := \Harnessing, 9 ; st.back()$, which refines $s = Back \to s := \Harnessing ; st.back()$ under~$R_3$. The conditions $(\bf N_2)$ and $(\bf N_3)$ are shown similarly.

\subsection*{Refinement 4: Introducing Santa's Shop}

\newcommand{\Welcoming}{\textit{Welcoming}}
\newcommand{\Consulting}{\textit{Consulting}}
\newcommand{\Shop}{\textit{Shop}}

Class $\Santa_4$ splits the $\Helping$ state of $\Santa_2$ into $\Welcoming$ and $\Consulting$; field $p$ is the number of puzzled elves. Santa will be woken up only by a group of three elves but then has to consult each individually. The shop is here an active object that represents the collective behaviour of elves: class $\Shop_4$ maintains a count of the number of elves of the current group that still have to consult with Santa:

\begin{lstlisting}
class Santa$_4$
    var s: {Sleeping, Harnessing, Riding, Welcoming, Consulting} = Sleeping
    var b: boolean = false
    var p: 0 .. 3 = 0
    method back()
        b := true
    method harness()
        s = Harnessing $\to$ s := Riding
    method pull()
        s = Riding $\to$ s, b := Sleeping, false
    method puzzled()
        p := 3
    method enter()
        s = Welcoming $\to$ s := Consulting
    method consult()
        s = Consulting $\to$ p := p - 1 ; if p > 0 then s := Welcoming else s := Sleeping
    action s = Sleeping $\land$ b $\to$ s := Harnessing
    action s = Sleeping $\land p = 3 \land \neg$b $\to$ s := Welcoming

class Shop$_4$(st: Santa$_4$)
    var s: {Puzzled, Entering, Consulting} = Puzzled
    var c: 0 .. 3 = 0
    action s = Puzzled $\to$ s, c := Entering, 3 ; st.puzzled()
    action s = Entering $\to$ s := Consulting ; st.enter()
    action s = Consulting $\to$ c := c - 1 ; if c > 0 then s := Entering else s:= Puzzled ; st.consult()
\end{lstlisting}

\noindent One Santa and one shop are created:

\begin{lstlisting}
st := new Santa$_4$ ; sh := new Shop$_4$(st)
\end{lstlisting}

\noindent As the coupling relation between $\Santa_2$ and $\Santa_4$ we take:

\begin{itemize}
    \item[] $R_4\, (s_2, b_2)\, (s_4, b_4, p_4) \defeq
    \begin{array}[t]{@{}l}
         (s_2 = s_4 \lor (s_2 = \Helping \land s_4 \in \{\Welcoming, \Consulting\})) \land \mbox{} \\
         b_2 = b_4
    \end{array}$
\end{itemize}

\noindent Class $\Sleigh_4$ refines $\Sleigh_2$ as the methods {\em back}, {\em harness}, {\em pull} refine themselves under $R_4$, new methods {\em puzzled}, {\em enter} stutter under $R_4$, new method {\em consult} stutters if $p > 0$ initially and refines $s_2 = \Helping \to s_2 := \Sleeping$ if $p = 0$ initially, and the two actions of $\Sleigh_4$ refine actions of $\Sleigh_2$. Formally, the conditions are:
\begin{itemize}
    \item[($\bf I$)] $R_4\,(\Sleeping, \False)\,(\Sleeping, \False, 0)$
    \item[($\bf M_1$)] $\Sleigh_2.back \sqsubseteq_{R_4} \Sleigh_4.back$
    \item[($\bf M_2$)] $\Sleigh_2.harness \sqsubseteq_{R_4} \Sleigh_4.harness$
    \item[($\bf M_3$)] $\Sleigh_2.pull \sqsubseteq_{R_4} \Sleigh_4.pull$
    \item[($\bf N_1$)] $\Skip \sqsubseteq_{R_4} \Sleigh_4.puzzled$
    \item[($\bf N_2$)] $\Skip \sqsubseteq_{R_4} \Sleigh_4.enter$
    \item[($\bf N_3$)] $\Skip \sqsubseteq_{R_4} \Sleigh_4.consult$ ~~or~~ $s_2 = \Helping \to s_2 := \Sleeping \sqsubseteq_{R_4} \Sleigh_4.consult$
    \item[($\bf A_1$)] $s_2 = \Sleeping \land b_2 \to s_2 := \Harnessing \sqsubseteq_{R_4} s_4 = \Sleeping \land b_4 \to s_4 := \Harnessing$
    \item[($\bf A_2$)] $s_2 = \Sleeping \land \neg b_2 \to s_2 := \Helping \sqsubseteq_{R_4} s_4 = \Sleeping \land p_4 = 3 \land \neg b_4 \to s_4 := \Welcoming$
\end{itemize}

\subsection*{Refinement 5: Introducing Elves}

This step leaves Santa, the sleigh, and the reindeer unchanged, refines the shop into a passive shop, and introduced active elves.

\begin{lstlisting}
class Shop$_5$(st: Santa)
    var s: {Puzzled, Entering, Consulting} = Puzzled
    var c: 0 .. 3 = 0
    method puzzled()
        s = Puzzled $\to$ c := c + 1 ; if c = 3 then (s := Entering; st.puzzled())
    method enter()
        s = Entering $\to$ s := Consulting ; st.enter()
    method consult()
        s = Consulting $\to$ c := c - 1 ; if c > 0 then s := Entering else s:= Puzzled ; st.consult()

class Elf$_5$(sh: Shop)
    action sh.puzzled() ; sh.enter() ;  sh.consult()
\end{lstlisting}

\newcommand{\Puzzled}{\textit{Puzzled}}
\newcommand{\Entering}{\textit{Entering}}
\newcommand{\If}{\textbf{if }}
\newcommand{\Then}{{\textbf{ then }}}
\newcommand{\Else}{{\textbf{ else }}}

\noindent One shop and 20 elves are created:

\begin{lstlisting}
sh := new Shop$_5$ ; for i := 1 to 20 do new Elf$_5$(sh)
\end{lstlisting}

\noindent The coupling relation between $\Shop_4$ and $\Shop_5$ includes the identity relation on $s$. The count $c$ is also identical except in state \Puzzled, as in $\Shop_5$ the elves may increment $c$ one by one but in $\Shop_4$ it is set to $3$ at once:

\begin{itemize}
    \item[] $R_5\, (s_4, c_4)\, (s_5, c_5) \defeq s_4 = s_5 \land (c_4 = c_5 \lor (s_5 = Puzzled \land c_5 < 3))$
\end{itemize}

\noindent Class $\Shop_5$ refines $\Shop_4$ as the new method {\em puzzled} stutters under $R_5$ if $c_5 < 2$ and refines the action $s_4 = \Puzzled \to s_4, c_4 := \Entering, 3 ; st.puzzled()$ if $c_5 = 2$, new method {\em enter} refines $s_4 = \Entering \to s_4 := \Consulting ; st.enter()$, and new method {\em consult} refines $s_4 = \Consulting \to c_4 := c_4 - 1 ; \If c_4 > 0 \Then s_4 := \Entering \Else s_4 := \Puzzled ; st.consult()$. Formally, the conditions are:

\begin{itemize}
    \item[($\bf I$)] $R_5\,(\Puzzled, 0)\,(\Puzzled, 0)$
    \item[($\bf N_1$)] $\Skip \sqsubseteq_{R_5} \Shop_5.puzzled$ ~~or~~ $s_4 = \Puzzled \to s_4, c_4 := \Entering, 3 ; st.puzzled() \sqsubseteq_{R_5} \Shop_5.puzzled$
    \item[($\bf N_2$)] $s_4 = \Entering \to s_4 := \Consulting ; st.enter() \sqsubseteq_{R_5} \Shop_5.enter$
    \item[($\bf N_3$)] $s_4 = \Consulting \to c_4 := c_4 - 1 ; \If c_4 > 0 \Then s_4 := \Entering \Else s_4 := \Puzzled ; st.consult() \sqsubseteq_{R_5} $\\
    \qquad $\Shop_5.consult$
\end{itemize}

\subsection*{Summary of Refinement Steps}

The final versions of classes \Santa, \Sleigh, {\em Reindeer}, \Shop, and {\em Elf} are:

\begin{lstlisting}
class Santa
    var s: {Sleeping, Harnessing, Riding, Welcoming, Consulting} = Sleeping
    var b: boolean = false
    var p: 0 .. 3 = 0
    method back()
        b := true
    method harness()
        s = Harnessing $\to$ s := Riding
    method pull()
        s = Riding $\to$ s, b := Sleeping, false
    method puzzled()
        p := 3
    method enter()
        s = Welcoming $\to$ s := Consulting
    method consult()
        s = Consulting $\to$ p := p - 1 ; if p > 0 then s := Welcoming else s := Sleeping
    action s = Sleeping $\land$ b $\to$ s := Harnessing
    action s = Sleeping $\land p = 3 \land \neg$ b$\to$ s := Welcoming

class Sleigh(st: Santa)
    var s: {Back, Harnessing, Pulling} = Back
    var c: 0 .. 9 = 9
    method back()
        s = Back $\to$ c := c - 1 ; if c = 0 then (s, c := Harnessing, 9 ; st.back())
    method harness()
        s = Harnessing $\to$ c := c - 1 ; if c = 0 then (s, c := Pulling, 9 ; st.harness())
    method pull()
        s = Pulling $\to$ c := c - 1 ; if c = 0 then (s, c := Back, 9 ; st.pull())

class Reindeer(sl: Sleigh)
    action sl.back() ; sl.harness() ; sl.pull()

class Shop(st: Santa)
    var s: {Puzzled, Entering, Consulting} = Puzzled
    var c: 0 .. 3 = 0
    method puzzled()
        s = Puzzled $\to$ c := c + 1 ; if c = 3 then (s := Entering; st.puzzled())
    method enter()
        s = Entering $\to$ s := Consulting ; st.enter()
    method consult()
        s = Consulting $\to$ c := c - 1 ; if c > 0 then s := Entering else s:= Puzzled ; st.consult()

class Elf(sh: Shop)
    action sh.puzzled() ; sh.enter() ;  sh.consult()
\end{lstlisting}

The main program creates active objects for Santa, reindeer, and elves; these use the passive sleigh and shop objects for synchronization:

\begin{lstlisting}
st := new Santa
sl := new Sleigh(st) ; for i := 1 to 9 do new Reindeer(sl)
sh := new Shop(st) ; for i := 1 to 20 do new Elf(sh)
\end{lstlisting}

\section{Results}

We have implemented an experimental compiler for Lime, a language that closely follows the above theory of concurrent objects. Appendix A contains the Lime implementation of the Santa Claus Problem. The key contributions of the compiler are the management of dynamically growing stacks, the efficient evaluation of method and action guards, a mapping of actions to coroutines, and a distribution of coroutines onto processor cores. The details are in~\cite{Yao18}. 

The Lime implementation is compared to implementations in C using semaphores of the Pthreads library, in Go using channels, and in Java using monitors, see Appendix A. Table~\ref{result} shows the running times for Santa with 9 reindeer and 20 elves. Santa's division of work is that for 10,000 rounds until retirement, he rides the sleigh 2,000 times and helps 8,000 times groups of three elves, or for 20 elves, each elf on average 1,200 times. For 100,000 and 1,000,000 rounds until Santa's retirement the ratio is the same. Some observations are in order:

\begin{table}\centering
\begin{tabular}{ |p{1.8cm}||p{2.7cm} | p{2.9cm}|p{2.7cm}| p{3.1cm}| }
 \hline
 %\multicolumn{4}{|c|}{Execution time (s)} \\
 %\hline
 % Lime 0.03695/0.0359/0.00255 297.9/295.15/3.8  2911.15/2903.8/7.9
Repetitions of Santa & Lime (guards) &C (semaphores) &Go (channels) &Java (monitors) \\
 \hline
 10,000    & 0.04 / 0.04 / 0.00   & 0.87 / 0.26 / 1.18 &   0.08 / 0.12 / 0.01 & 6.38 / 2.48 / 5.30\\
 
 100,000  & 0.30 / 0.30 / 0.00  & 8.82 / 2.50 / 12.0  & 0.77 / 1.18 / 0.06  & 60.3 / 21.6 / 52.0\\
 1,000,000 & 2.91 / 2.90 / 0.01 & 93.0 / 24.8 / 123 &  7.51 / 11.6 / 0.55 & $\approx$ 534 / 159 / 509 \\
 \hline
\end{tabular}
\caption{Execution time in sec on AMD Ryzen Threadripper 1950X 16 core (32 threads) processor with 32 GB memory under Ubuntu 16.04. The compilers used are gcc 5.4.0, Java 9.0.4, Go 1.8.3 linux/amd64. The times are reported as the average real / user / system times of 20 runs. Only a single run was used for Java with 1,000,000 repetitions of Santa.}
\label{result}
\end{table}

\begin{itemize}
    \item The Java implementation uses a single monitor for all synchronization. While it would be natural to have Santa, reindeer, and elf processes as well as sleigh, shop, and Santa monitors (synchronizing reindeer, elves, and the sleigh / shop, respectively), this leads to the nested monitor call problem, for example when elves are calling the shop and the shop calls Santa. Ben-Ari's and our implementation use, therefore, a single monitor with the functionality of sleigh, shop, and Santa monitors. This limits concurrency, e.g. reindeer and elves cannot assemble independently. Java necessitates that each monitor method contains a {\em notifyAll$()$} for waking up all threads, most of which will immediately sleep again. The timing results confirm that this is wasteful; in particular, the ratio between user and system times make the synchronization effort evident.
    \item The C implementation uses operating systems threads, which require more cycles when switching than lightweight threads as used by Lime, Go, and Java. Compared to Java with monitors, only the ``right'' threads are woken up, but the ratio of user to system time tells that switching operating systems threads is expensive.
    \item The Go implementation uses CSP-like synchronous channels, which are particularly suitable for barrier synchronization with Santa; by comparison, of the semaphore $P()$ and $V()$ operations, only one blocks, meaning that two semaphores are needed for each synchronization point. The goroutines (lightweight threads) of Go are mapped to coroutines, like in Lime, and distributed over cores (like in Lime), leading to good performance. Go does not support priorities when receiving or sending over channels, so to give reindeer priority over elves, a workaround is needed.
    \item The Lime runtime system is designed for very quickly switching between actions when a guard blocks. Since the bodies of methods and actions in the Santa Claus Problem are short, this pays off. Interestingly, the real time is the user time, suggesting that only one core was active. The Lime runtime system is also designed for distributing a very large number of concurrent objects among cores. As there are relatively few objects here and the bodies of methods are so short that work stealing is not effective, the Lime runtime system is not able to utilize more than one core.
\end{itemize}

The Haskell implementation of Peyton Jones was not included as its proper functioning depends on the presence of delay statements. Trono's implementation does not run reliably under Pthreads and has more relaxed synchronization constraints than the Lime version, so is not included in the comparison either.
% - 9 reindeer and 20 elves
% - each elf participates 12,000 times, so Santa helps elves 12,000 * 20 elves / 3 elves per round = 80,000 times
% - each reindeer participates 20,000 times, so Santa delivers 20,000 * 9 reindeer / 9 reindeer per ride = 20,000 times
% - Santa works 100,000 times

\section{Discussion}

In ongoing work, we observed on a number of concurrency examples, that Lime compares favourably to all other languages that we compared with~\cite{Yao18}, which made us wonder if that would be the case for the Santa Claus Problem as well. It took us by surprise that Lime is close to three times faster than Go, about 32 times faster than C, and more than 180 times faster than Java when measuring elapsed time. This line of work provides evidence that the evaluation of guards in methods and actions, compared to synchronizing with semaphores and monitors or sending over channels, is not intrinsically less efficient; the overall efficiency depends more on the techniques used for mapping actions to coroutines and quickly switching between them. This is encouraging for the use of verification and refinement techniques that rely on guards, as these can an applied to highly efficient implementations.

\nocite{*}
\bibliographystyle{eptcs}
\bibliography{santa}

\section*{Appendix A}
These implementations are used in the comparison of timing results.

\lstset{
    basicstyle=\small,
    captionpos=t,
    columns=flexible,
	morekeywords={class, method, action, init, for, if, do, when, then, var, int, boolean},
	deletekeywords={this}}

\begin{lstlisting}[basicstyle=\footnotesize, columns=flexible,caption=Implementation with Lime]
class Santa
    var s: {Sleeping, Harnessing, Riding, Welcoming, Consulting}
    var b: boolean
    var p: int
    init()
        s, b, p := Sleeping, false, 0
    method back()
        b := true
    method harness()
        when s = Harnessing do 
            s := Riding
    method pull()
        when s = Riding do
            s, b := Sleeping, false
    method puzzled()
        p := 3
    method enter()
        when s = Welcoming do 
            s := Consulting
    method consult()
        when s = Consulting do
            p := p - 1
            if p > 0 then 
                s := Welcoming 
            else 
                s := Sleeping
    action action1
        when s = Sleeping and b do 
            s := Harnessing
    action action2
        when s = Sleeping and p = 3 and not b do 
            s := Welcoming

class Sleigh
    var s: {Back, Harnessing, Pulling}
    var c: int
    var st: Santa
    init(santa: Santa)
        s, c, st := Back, 9, santa
    method back()
        when s = Back do
            c := c - 1 
            if c = 0 then 
                s, c := Harnessing, 9
                st.back() 
    method harness()
        when s = Harnessing do 
            c := c - 1 
            if c = 0 then 
                s, c := Pulling, 9
                st.harness() 
    method pull()
        when s = Pulling do
            c := c - 1  
            if c = 0 then 
                s, c := Back, 9
                st.pull()

class Reindeer
    var sl: Sleigh
    init (sleigh: Sleigh)
        sl := sleigh
    action action1
        sl.back()
        sl.harness()
        sl.pull()

class Shop
    var s: {Puzzled, Entering, Consulting}
    var c: int
    init(santa: Santa)
        s, c, st := Puzzled, 0, santa
    method puzzled()
        when s = Puzzled do  
            c := c + 1 
            if c = 3 then 
                s := Entering
                st.puzzled() 
    method enter()
        when s = Entering do
            s := Consulting
            st.enter()
    method consult()
        when s = Consulting do
            c := c - 1 
            if c > 0 then 
                s := Entering
            else
                s := Puzzled
            st.consult()
            
class Elf
    var sh: Shop
    init(shop: Shop)
        sh := shop
    action action1
        sh.puzzled()
        sh.enter()  
        sh.consult()
        
class Start
    var st: Santa
    var sl: Sleigh
    var sh: Shop
    init()
        st := new Santa()  
        sl := new Sleigh(st)  
        sh := new Shop(st)
        for i := 1 to 9 do new Reindeer(sl)
        for i := 1 to 20 do new Elf(sh)
\end{lstlisting}

\begin{lstlisting}[language=C,basicstyle=\footnotesize,columns=flexible,caption=Implementation with C]
#include <stdbool.h>
#include <pthread.h>
#include <semaphore.h>
#define P(sem)  (sem_wait(&(sem)))  /* uses P and V for the wait and ... */
#define V(sem)  (sem_post(&(sem)))  /* ... signal semaphore operations */

sem_t wakeup, wakeupReindeer, wakeupElves;
sem_t harness, harnessDone;
sem_t pull, pullDone;
sem_t enter, enterDone;
sem_t consult, consultDone;
sem_t reindeerBack, reindeerBackDone;
sem_t reindeerHarness, reindeerHarnessDone;
sem_t reindeerPull, reindeerPullDone;
sem_t elfPuzzled, elfPuzzledDone;
sem_t elfEnter, elfEnterDone;
sem_t elfConsult, elfConsultDone;

bool b;

void *Santa(void *arg) {
    for (int t = 0; t < 10000; t++) { // Sleeping
        P(wakeup); // woken up by Sleigh or Shop
        if (b) { // Delivering
            b = false; V(wakeupReindeer);
            P(harness); V(harnessDone);
            P(pull); V(pullDone);
        } else { // Helping
            V(wakeupElves);
            for (int i = 0; i < 3; i++) {
                P(enter); V(enterDone);
                P(consult); V(consultDone);
            }
        }
    }
}
void *Sleigh(void *arg) {
    for (;;) {
        for (int i = 0; i < 9; i++) V(reindeerBack);
        for (int i = 0; i < 9; i++) P(reindeerBackDone);
        b = true; V(wakeup); P(wakeupReindeer);
        for (int i = 0; i < 9; i++) V(reindeerHarness);
        for (int i = 0; i < 9; i++) P(reindeerHarnessDone);
        V(harness); P(harnessDone);
        for (int i = 0; i < 9; i ++) V(reindeerPull);
        for (int i = 0; i < 9; i ++) P(reindeerPullDone);
        V(pull); P(pullDone);
    }
}
void *Reindeer(void *arg) {
    for (int t = 0; t < 2000; t++) {
        P(reindeerBack); V(reindeerBackDone);
        P(reindeerHarness); V(reindeerHarnessDone);
        P(reindeerPull); V(reindeerPullDone);
    }
}
void *Shop(void *arg) {
    for (;;) {
        for (int i = 0; i < 3; i++) V(elfPuzzled);
        for (int i = 0; i < 3; i++) P(elfPuzzledDone);
        V(wakeup); P(wakeupElves);
        for (int i = 0; i < 3; i++) {
            V(elfEnter); P(elfEnterDone);
            V(enter); P(enterDone);
            V(elfConsult); P(elfConsultDone);
            V(consult); P(consultDone);
        }
    }
}
void *Elf(void *arg) {
    for (;;) {
        P(elfPuzzled); V(elfPuzzledDone);
        P(elfEnter); V(elfEnterDone);
        P(elfConsult); V(elfConsultDone);
    }
}
void main() {
    sem_init(&wakeup, 0, 0); sem_init(&wakeupReindeer, 0, 0); sem_init(&wakeupElves, 0, 0);
    sem_init(&harness, 0, 0); sem_init(&harnessDone, 0, 0);
    sem_init(&pull, 0, 0); sem_init(&pullDone, 0, 0);
    sem_init(&enter, 0, 0); sem_init(&enterDone, 0, 0);
    sem_init(&consult, 0, 0); sem_init(&consultDone, 0, 0);
    sem_init(&reindeerBack, 0, 0); sem_init(&reindeerBackDone, 0, 0);
    sem_init(&reindeerHarness, 0, 0); sem_init(&reindeerHarnessDone, 0, 0);
    sem_init(&reindeerPull, 0, 0); sem_init(&reindeerPullDone, 0, 0);
    sem_init(&elfPuzzled, 0, 0); sem_init(&elfPuzzledDone, 0, 0);
    sem_init(&elfEnter, 0, 0); sem_init(&elfEnterDone, 0, 0);
    sem_init(&elfConsult, 0, 0); sem_init(&elfConsultDone, 0, 0);
    pthread_t tid;
    for (int i = 0; i < 9; i++) pthread_create(&tid, NULL, Reindeer, NULL);
    for (int i = 0; i < 20; i++) pthread_create(&tid, NULL, Elf, NULL);
    pthread_create(&tid, NULL, Sleigh, NULL); pthread_create(&tid, NULL, Shop, NULL);
    pthread_create(&tid, NULL, Santa, NULL); pthread_join(tid, NULL);
}
\end{lstlisting}

\begin{lstlisting}[language=C,basicstyle=\footnotesize,columns=flexible,morekeywords={func,var,select,go,package},caption=Implementation with Go]
package main

var reindeerBack, reindeerHarness, reindeerPull chan bool
var back, harness, pull chan bool
var elfPuzzled, elfEnter, elfConsult chan bool
var puzzled, enter, consult chan bool
var done chan bool

func Santa() {
    b, p := false, false // reindeer back, elves puzzled
    for t := 0; t < 10000; t++ { // invariant: !b
        if !p { // neither reindeer back nor elves puzzled
            select { // wait for either one
            case <- back: b = true
            case <- puzzled: p = true
            }
        }
        if p { // elves puzzled
            select { // check if reindeer back as well
            case <- back: b = true
            default:
            }
        }
        // either b or p is true, pick one
        if b { // prefer reindeer
            <- harness ; <- pull ; b = false
        } else { // otherwise elves
            for i := 0; i < 3; i++ {
                <- enter ; <- consult
            }
            p = false
        }
    }
    done <- true
}
func Sleigh() {
    for {
        for i := 0; i < 9; i++ {<- reindeerBack}
        back <- true
        for i := 0; i < 9; i++ {<- reindeerHarness}
        harness <- true
        for i := 0; i < 9; i++ {<- reindeerPull}
        pull <- true
    }
}
func Shop() {
    for {
        for i := 0; i < 3; i++ {<- elfPuzzled}
        puzzled <- true
        for i := 0; i < 3; i++ {
            <- elfEnter ; enter <- true ; <- elfConsult ; consult <- true
        }
    }
}
func Reindeer() {
    for r := 0; r < 2000; r++ {
        reindeerBack <- true ; reindeerHarness <- true ; reindeerPull <- true
    }
}
func Elf() {
    for {
        elfPuzzled <- true ; elfEnter <- true ; elfConsult <- true
    }
}
func main() {
    reindeerBack, reindeerHarness, reindeerPull = make(chan bool), make(chan bool), make(chan bool)
    back, harness, pull = make(chan bool), make(chan bool), make(chan bool)
    elfPuzzled, elfEnter, elfConsult = make(chan bool), make(chan bool), make(chan bool)
    puzzled, enter, consult = make(chan bool), make(chan bool), make(chan bool)
    done = make(chan bool)
    go Santa(); go Sleigh(); go Shop()
    for i := 0; i < 9; i++ {go Reindeer()}
    for i := 0; i < 20; i++ {go Elf()}
    <- done
}
\end{lstlisting}

\begin{lstlisting}[language=Java,morekeywords={enum}, basicstyle=\footnotesize,columns=flexible,caption=Implementation with Java]
enum R {Relaxing, Back, Harnessing, Harnessed, Pulling, Done}
enum E {Working, Puzzled, Entering, Entered, Consulting, Enlightened}
enum Task {deliver, help}
class SantasShop {
    int rc = 9, ec = 3; // reindeer count, elf count
    R rs = R.Relaxing;  // state of reindeer
    E es = E.Working;   // state of elves

    synchronized void back() /* called by reindeer */ {
        while (rs != R.Relaxing) try {wait();} catch (Exception x) {} 
        rc -= 1; if (rc == 0) {rs = R.Back; rc = 9;} notifyAll();
    }
    synchronized void harness() /* called by reindeer */ {
        while (rs != R.Harnessing) try {wait();} catch (Exception x) {} 
        rc -= 1; if (rc == 0) {rs = R.Harnessed; rc = 9;} notifyAll();
    }
    synchronized void pull() /* called by reindeer */ {
        while (rs != R.Pulling) try {wait();} catch (Exception x) {} 
        rc -= 1; if (rc == 0) {rs = R.Done; rc = 9;} notifyAll();
    }

    synchronized void puzzled() /* called by elves */ {
        while (es != E.Working) try {wait();} catch (Exception x) {} 
        ec -= 1; if (ec == 0) {es = E.Puzzled; ec = 3;} notifyAll();
    }
    synchronized void enter() /* called by elves */ {
        while (es != E.Entering) try {wait();} catch (Exception x) {} 
        es = E.Entered; notifyAll();
    }
    synchronized void consult() /* called by elves */ {
        while (es != E.Consulting) try {wait();} catch (Exception x) {} 
        es = E.Enlightened; notifyAll();
    }
    
    synchronized Task wakeup() /* called by Santa */ { 
        while (rs != R.Back && es != E.Puzzled) try {wait();} catch (Exception x) {}
        if (rs == R.Back) {rs = R. Harnessing; notifyAll(); return Task.deliver;}
        else {es = E.Entering; notifyAll(); return Task.help;}
    }
    synchronized void hitch() /* called by Santa */ {
        while (rs != R.Harnessed) try {wait();} catch (Exception x) {} 
        rs = R.Pulling; notifyAll();
    }
    synchronized void ride() /* called by Santa */ {
        while (rs != R.Done) try {wait();} catch (Exception x) {} 
        rs = R.Relaxing; notifyAll();
    }
    synchronized void welcome() /* called by Santa */ {
        while (es != E.Entered) try {wait();} catch (Exception x) {}
        es = E.Consulting; notifyAll();
    }
    synchronized void explain() /* called by Santa */ {
        while (es != E.Enlightened) try {wait();} catch (Exception x) {}
        ec -= 1; if (ec == 0) {es = E.Working; ec = 3;} else es = E.Entering;
        notifyAll();
    }

    public static void main(String[] args) {
        SantasShop shop = new SantasShop();
        new Santa(shop).start();
        for (int i = 0; i < 9; i++) new Reindeer(shop).start();
        for (int i = 0; i < 20; i++) {Thread e = new Elf(shop, i); e.setDaemon(true); e.start();}
    }
}
class Santa extends Thread {
    SantasShop shop;
    Santa(SantasShop ss) {shop = ss;}
    public void run() {
        for (int t = 0; t < 10000; t++) {
            Task task = shop.wakeup();
            if (task == Task.deliver) {
                shop.hitch(); shop.ride();
            } else {
                for (int i = 0; i < 3; i++) {shop.welcome(); shop.explain();}
            }
        }
    }
}
class Reindeer extends Thread {
    SantasShop shop;
    Reindeer(SantasShop ss) {shop = ss;}
    public void run() {
        for (int t = 0; t < 2000; t++) {shop.back(); shop.harness(); shop.pull();}
    }
}
class Elf extends Thread {
    SantasShop shop; int num;
    Elf(SantasShop ss, int n) {shop = ss; num = n;}
    public void run() {
        for (;;) {shop.puzzled(); shop.enter(); shop.consult();}
    }
}
\end{lstlisting}

\end{document}